\documentstyle [12pt]{article}
\begin{document}
\begin{titlepage}
\hspace{9cm} ULB--PMIF--92/04

\vspace{3cm}
\begin{centering}

{\Large BRST-anti-BRST Antifield Formalism :
The Example of the Freedman-Townsend Model}
\vspace{1cm}

{\large G. Barnich$^*$, R. Constantinescu$^\dagger$, and P.
Grgoire$^\ddagger$}\\

Facult\'e des Sciences, Universit\'e Libre de Bruxelles,\\
Campus Plaine C.P. 231, B-1050 Bruxelles, Belgium\\
\vspace{2cm}
{\large Abstract}

\end{centering}
\vspace{.5cm}

The general BRST-anti-BRST construction in the framework of the
antifield-antibracket
formalism is illustrated in the case of the Freedmann-Townsend
model.
\vspace{3cm}

\noindent
{\footnotesize
($^*$)Aspirant au Fonds National de la Recherche Scientifique,
Belgique. \\
($^\dagger$)Permanent address at Department of Theoretical
Physics, University of
Craiova, 13 rue A.I. Cuza, 1100 Craiova, Romania.\\
($^\ddagger$)Chercheur IRSIA, Belgique.}

\end{titlepage}
\pagebreak
\section{Introduction}
The most general method for quantizing gauge systems in a
manifestly covariant
manner is the antifield-antibracket formalism
\cite{BatVil1,BatVil2,BatVil3,FisHen1}.
This method allows to express the BRST symmetry in the
Lagrangian context.
Immediatly after the discovery of this symmetry, the anti-BRST
transformation
was formulated for Yang-Mills models. Recently, there has been
an increasing
interest in a systematic formulation of the BRST-anti-BRST
symmetry in both
the  Hamiltonian \cite{BatLav1,BatLav2,GreHen1,GreHen2} and the
Lagrangian
\cite{BatLav3,BatLav4,Hen} context. Those methods rely on the
construction
of two nilpotent anticommuting operators $ s_1 $ and $ s_2 $
leaving the action
invariant.

A model that has been intensively studied from the Lagrangian
point of view is the
Freedman-Townsend model. Its main interest lies in the fact
that ({\em i}) it is
equivalent to the non-linear $ \sigma $-model \cite{FreTow},
({\em ii}) the algebraic structure
of its gauge symmetries is similar to that of Witten's string
theory \cite{Wit}, ({\em iii})
even if its BRST structure is well understood
\cite{AlvGri,BatGom,BauBer}, the attemps to incorporate the
anti-BRST symmetry have not been entirely satisfactory
\cite{BauThi}. Our purpose in
this article is to illustrate the general Lagrangian BRST-anti-
BRST method in the case
of this representative model.

Our starting point is the classical action ( where the trace
over group indices is understood)
\begin{equation}
S_{0}[B_{\mu\nu}(x),A_{\mu}(x)]
=\int d^{4}x\; \Bigg\{ \frac{1}{4} \epsilon^{\mu\nu\rho\sigma}
B_{\mu\nu}(x)F_{\rho\sigma}(x) + \frac{1}{2}
A_{\mu}(x)A^{\mu}(x) \Bigg\},
\end{equation}
which is invariant under the gauge transformations
\begin{equation}
\delta_{g}B_{\mu\nu}(x)=\nabla_{[\mu}\xi_{\nu]}(x) \; ,
\; \delta_{g}A_{\mu}=0. \label{gaugetransf}
\end{equation}
The bosonic fields $ A_{\mu} $ and $ B_{\mu\nu} $ take values
in some semi-simple compact
Lie algebra $ {\cal A} $.
The gauge transformations (\ref{gaugetransf}) are not all
independent; indeed,
it is easy to see that if one takes for gauge parameters
$ \xi_{\mu}(x)=\nabla_{\mu}\xi(x) $, then the gauge
transformations reduce to an
antisymmetric combination of the equations of motion :
\begin {equation}
\delta_{g}B_{\mu\nu}(x)=-\frac{1}{2}\epsilon_{\mu\nu\rho\sigma}
[\frac{\delta S_{0}}{\delta B_{\rho\sigma}(x)},\xi(x)].
\end{equation}
In the DeWitt notations, the gauge generators are written as
\begin{equation}
R^{a\;\rho}_{\mu\nu
b}(x,y)=\delta^{\rho}_{[\nu}(\nabla_{x^{\mu}]})^{a}_{b}
\delta^{4}(x-y),
\end{equation}
while the first order reducibility functions are given by
\begin{equation}
Z^{b}_{\rho c}(y,z)=(\nabla_{y^{\rho}})^{b}_{c}\delta^{4}(y-z).
\end{equation}
In these notations, the reducibility relation takes the form
\begin{equation}
\int d^{4}y R^{a\;\rho}_{\mu\nu b}(x,y)Z^{b}_{\rho c}(y,z)=
-\frac{1}{2}\epsilon_{\mu\nu\rho\sigma}\frac{\delta
S_{0}}{\delta B_{b\rho\sigma}(x)}
f^{a}_{bc}\delta^{4}(x-z), \label{redrel}
\end{equation}
where the $f^{a}_{bc}$'s are totally antisymmetric structure
constants of the
Lie algebra $ {\cal A} $.
The $Z$'s are a complete and irreducible set of first order
reducibility coefficients,
that is, with this choice of $R$ and $Z$ the Freedman-Townsend
model is a first order reducible
theory. The particularity of this description is that
reducibility
only holds on-shell. In the usual BRST antifield-antibracket
formalism this implies that the
solution of the master equation contains supplementary terms in
comparison with
off-shell reducible gauge systems. This will also be true for
the BRST-anti-BRST
Lagrangian treatment.

\section{The BRST-anti-BRST Lagrangian formalism for a first
order reducible
gauge system}
{}From a generic point of view, the Freedman-Townsend model may
be characterized
by an action $ S_{0}[q^{i}] $ where the $ q^{i} $ denote all
the fields ; in our case one has the
correspondance:
\begin{equation}
q^{i} \longleftrightarrow
(A_{\mu}(x),B_{\mu\nu}(x)).\label{dic1}
\end{equation}
\pagebreak

\noindent In particular, here the $q^{i}$'s are bosonic. The
action $ S_{0}[q^{i}] $ is invariant under some gauge
transformations
\begin{equation}
\delta_{g}q^{i} =R^{i}_{\alpha_{0}}\xi^{\alpha_{0}} ,
\label{gengaugetransf}
\end{equation}
which are abelian off-shell.
If $ G_{i}=0 $ denote the equations of motion, then the Noether
identities
associated to the gauge symmetries (\ref{gengaugetransf}) are
\begin{equation}
G_{i}R^{i}_{\alpha_{0}}=0.
\end{equation}
The first order reducibility relations are given by
\begin{equation}
R^{i}_{\alpha_{0}}Z^{\alpha_{0}}_{\alpha_{1}} =
G_{j}T^{ji}_{\alpha_{1}}.
\end{equation}
For the Freedman-Townsend model, one has the following
correspondance
\begin{equation}
R^{i}_{\alpha_{0}} \longleftrightarrow \left( \begin{array}{c}
0 \\
R^{a\;\rho}_{\mu\nu
b}(x,y)=\delta^{\rho}_{[\nu}(\nabla_{x^{\mu}]})^{a}_{b}
\delta^{4}(x-y) \end{array} \right) , \label{dic2}
\end{equation}
\vskip 0.2cm
\begin{equation}
Z^{\alpha_{0}}_{\alpha_{1}} \longleftrightarrow
Z^{b}_{\rho c}(y,z)=(\nabla_{y^{\rho}})^{b}_{c}\delta^{4}(y-z)
, \label{dic3}
\end{equation}
\vskip 0.2cm
\begin{equation}
T^{ij}_{\alpha_{1}} \longleftrightarrow
\left( \begin{array}{cc}
0 & 0 \\
0 & {\displaystyle
-\frac{1}{2}\epsilon_{\mu\nu\rho\sigma}
f^{a}_{bc}\delta^{4}(x-z)} \end{array} \right) . \label{corr3}
\end{equation}

\subsection{The ghost spectrum}
The main idea behind the BRST-anti-BRST algebraic structure
consists in ({\em i}) the
doubling of the initial gauge symmetries by introducing a
double set of initial ghosts;
and ({\em ii}) in the introduction of a bigrading called the
{\em new ghost bigrading}
and denoted by $ bingh $ \cite{GreHen1,GreHen2}. Thus, to the
gauge generators
and the first order reducibility functions
we associate the ghosts
\begin{equation}
R^{i}_{\alpha_{0}} \longrightarrow
(\stackrel{(1,0)}{\varphi^{\alpha_{0}}},
\stackrel{(0,1)}{\varphi^{\alpha_{0}}}) ,
\end{equation}
\begin{equation}
Z^{\alpha_{0}}_{\alpha_{1}} \longrightarrow
(\stackrel{(2,0)}{\varphi^{\alpha_{1}}},
\stackrel{(0,2)}{\varphi^{\alpha_{1}}}).
\end{equation}
The superscript $ (a,b) $ denotes the $ bingh $ of the
corresponding generators.
The total ghost spectrum is according to Batalin et al.
\cite{BatLav4} :
\begin{displaymath}
\stackrel{(1,0)}{\varphi^{\alpha_{0}}},
\stackrel{(0,1)}{\varphi^{\alpha_{0}}},
\stackrel{(1,1)}{\pi^{\alpha_{0}}},
\stackrel{(2,0)}{\varphi^{\alpha_{1}}},
\stackrel{(1,1)}{\varphi^{\alpha_{1}}},
\stackrel{(0,2)}{\varphi^{\alpha_{1}}},
\stackrel{(2,1)}{\pi^{\alpha_{1}}},
\stackrel{(1,2)}{\pi^{\alpha_{1}}}
\end{displaymath}
It can be understood by using the extended longitudinal
differential of \cite{HenTei,HenSta}.
Instead of the two sets of ghosts of the usual BRST formalism,
the additional
symmetry requires eight different sets of ghosts.

\subsection{The antifield spectrum and the boundary conditions}
There are two brackets in the BRST-anti-BRST formalism
\cite{BatLav3,BatLav4,Hen},
respectively denoted by $ (\cdot,\cdot)_{1} $ and $
(\cdot,\cdot)_{2} $. This
implies that to each field there will correspond two
antifields, one conjugated
in the first antibracket and the other conjugated in the second
antibracket.
This leads to the following field-antifield spectrum :
\begin{displaymath}
\begin{array}{ccc}
\Bigg| \begin{array}{c} \stackrel{(0,0)}{q^{i}} \\
\begin{array}{rr} \stackrel{(-1,0)*}{q_{i(1)}} &
\stackrel{(0,-1)*}{q_{i(2)}} \end{array} \end{array} & \Bigg|
\begin{array}{c}
\stackrel{(1,0)}{\varphi^{\alpha_{0}}} \\ \begin{array}{rr}
\stackrel{(-2,0)*}{\varphi_{\alpha_{0}(1)}} &
\stackrel{(-1,-1)*}{\varphi_{\alpha_{0}(2)}} \end{array}
\end{array} & \Bigg|
\begin{array}{c}
\stackrel{(0,1)}{\varphi^{\alpha_{0}}} \\ \begin{array}{rr}
\stackrel{(-1,-1)*}{\varphi_{\alpha_{0}(1)}} &
\stackrel{(0,-2)*}{\varphi_{\alpha_{0}(2)}} \end{array}
\end{array} \Bigg|
\end{array}
\end{displaymath}
\begin{displaymath}
\Bigg| \begin{array}{ccc}
\begin{array}{c}
\stackrel{(1,1)}{\pi^{\alpha_{0}}} \\ \begin{array}{rr}
\stackrel{(-2,-1)*}{\pi_{\alpha_{0}(1)}} &
\stackrel{(-1,-2)*}{\pi_{\alpha_{0}(2)}} \end{array}
\end{array} & \Bigg|
\begin{array}{c}
\stackrel{(2,0)}{\varphi^{\alpha_{1}}} \\ \begin{array}{rr}
\stackrel{(-3,0)*}{\varphi_{\alpha_{1}(1)}} &
\stackrel{(-2,-1)*}{\varphi_{\alpha_{1}(2)}} \end{array}
\end{array} & \Bigg|
\begin{array}{c}
\stackrel{(1,1)}{\varphi^{\alpha_{1}}} \\ \begin{array}{rr}
\stackrel{(-2,-1)*}{\varphi_{\alpha_{1}(1)}} &
\stackrel{(-1,-2)*}{\varphi_{\alpha_{1}(2)}} \end{array}
\end{array} \Bigg|
\end{array}
\end{displaymath}
\begin{displaymath}
\Bigg| \begin{array}{ccc}
\begin{array}{c}
\stackrel{(0,2)}{\varphi^{\alpha_{1}}} \\ \begin{array}{rr}
\stackrel{(-1,-2)*}{\varphi_{\alpha_{1}(1)}} &
\stackrel{(0,-3)*}{\varphi_{\alpha_{1}(2)}} \end{array}
\end{array} & \Bigg|
\begin{array}{c}
\stackrel{(2,1)}{\pi^{\alpha_{1}}} \\ \begin{array}{rr}
\stackrel{(-3,-1)*}{\pi_{\alpha_{1}(1)}} &
\stackrel{(-2,-2)*}{\pi_{\alpha_{1}(2)}} \end{array}
\end{array} & \Bigg|
\begin{array}{c}
\stackrel{(1,2)}{\pi^{\alpha_{1}}} \\ \begin{array}{rr}
\stackrel{(-2,-2)*}{\pi_{\alpha_{1}(1)}} &
\stackrel{(-1,-3)*}{\pi_{\alpha_{1}(2)}} \end{array}
\end{array} \Bigg|
\end{array}
\end{displaymath}
\vskip1cm
The resolution bidegree $bires=(res_{(1)},res_{(2)}) $ is given
by $ -bingh $
and is non-zero only for the antifields. The extended master
equation will be decomposed
according to the resolution degree
$ res=res_{(1)}+res_{(2)} $ . Following \cite{BatLav4,Hen}, we
introduce
supplementary antifields, referred to as the "bar-variables":
\begin{equation}
\stackrel{(-1,-1)}{\bar{q}_{i}}, \stackrel{(-2,-
1)}{\bar{\varphi}_{\alpha_{0}}},
\stackrel{(-1,-2)}{\bar{\varphi}_{\alpha_{0}}}, \stackrel{(-2,-
2)}{\bar{\pi}_{\alpha_{0}}},
\stackrel{(-3,-1)}{\bar{\varphi}_{\alpha_{1}}}, \stackrel{(-2,-
2)}{\bar{\varphi}_{\alpha_{1}}},
\stackrel{(-1,-3)}{\bar{\varphi}_{\alpha_{1}}}, \stackrel{(-3,-
2)}{\bar{\pi}_{\alpha_{1}}},
\stackrel{(-2,-3)}{\bar{\pi}_{\alpha_{1}}}.
\end{equation}
Having the complete fields-antifields spectrum, we can now give
the boundary conditions of the
solution of the extended master equation:
\begin{equation}
{\raise1.2pt\hbox{$ \stackrel{(0)}{S}$}}= S_{0} ,
\end{equation}
\begin{equation}
{\raise1.2pt\hbox{$ \stackrel{(1)}{S}$}} = \stackrel{(-
1,0)*}{q_{i(1)}}R^{i}_{\alpha_{0}}\stackrel{(1,0)}{
\varphi^{\alpha_{0}}} +\stackrel{(0,-
1)*}{q_{i(1)}}R^{i}_{\alpha_{0}}\stackrel{(0,1)}{
\varphi^{\alpha_{0}}},
\end{equation}
\begin{eqnarray}
{\raise1.2pt\hbox{$ \stackrel{(2)}{S}$}} =
\stackrel{(-
2,0)*}{\varphi_{\alpha_{0}(1)}}Z^{\alpha_{0}}_{\alpha_{1}}
\stackrel{(2,0)}{\varphi^{\alpha_{1}}} +
\stackrel{(0,-2)*}{\varphi_{\alpha_{0}(2)}}
Z^{\alpha_{0}}_{\alpha_{1}}\stackrel{(0,2)}{\varphi^{\alpha_{1}
}} +
{1\over2}(\stackrel{(-1,-1)*}{\varphi_{\alpha_{0}(1)}}+
\stackrel{(-1,-
1)*}{\varphi_{\alpha_{0}(2)}})Z^{\alpha_{0}}_{\alpha_{1}}
\stackrel{(1,1)}{\varphi^{\alpha_{1}}} \nonumber \\
+\left( \stackrel{(-1,-1)}{\bar{q}_{i}}R^{i}_{\alpha_{0}}
+(\stackrel{(-1,-1)*}{\varphi_{\alpha_{0}(2)}}-
\stackrel{(-1,-
1)*}{\varphi_{\alpha_{0}(1)}})R^{i}_{\alpha_{0}}\right)
\stackrel{(1,1)}{\pi^{\alpha_{0}}} + \ldots , \label{S2}
\end{eqnarray}
\begin{eqnarray}
{\raise1.2pt\hbox{$ \stackrel{(3)}{S}$}} = \left(
(\stackrel{(-2,-1)}{\bar{\varphi}_{\alpha_{0}}}- {1\over2}
\stackrel{(-2,-
1)*}{\pi_{\alpha_{0}(1)}})Z_{\alpha_{1}}^{\alpha_{0}}+
(\stackrel{(-2,-1)*}{\varphi_{\alpha_{1}(2)}}- \stackrel{(-2,-
1)*}{\varphi_{\alpha_{1}(1)}})\right)\stackrel{(2,1)}{
\pi^{\alpha_{1}}}
\nonumber \\
+\left( (\stackrel{(-1,-2)}{\bar{\varphi}_{\alpha_{0}}}- {1\over2}
\stackrel{(-1,-
2)*}{\pi_{\alpha_{0}(1)}})Z_{\alpha_{1}}^{\alpha_{0}}+
(\stackrel{(-1,-2)*}{\varphi_{\alpha_{1}(2)}}- \stackrel{(-1,-
2)*}{\varphi_{\alpha_{1}(1)}})\right)\stackrel{(1,2)}{
\pi^{\alpha_{1}}} + \ldots.
\end{eqnarray}

The total field-antifield spectrum as well as the boundary
conditions of the
master equation can be understood through homological arguments
\cite{GreHen1,GreHen2,Hen}.

\subsection{Resolution of the master equation}
In this section we will explicitly solve the classical
extended master equation
\begin{equation}
{1\over2}(S,S) + VS=0, \label{mastereq}
\end{equation}
for a first order on-shell reducible, off-shell abelian gauge
system\footnote
{The fact that the Freedman-Townsend model is abelian greatly
simplifies the
forthcoming computations. Nevertheless, the case of higher
order reducible gauge systems with
non-abelian open algebras can be treated exactly along the same
lines, the only difference being
the appearance of supplementary terms in the solution of the
master equation.}.
We introduce at this stage the generic notation $ \Phi^{A} $
for all the fields,
$ \Phi^{*}_{A(1)} $ and $ \Phi^{*}_{A(2)} $ for the antifields
respectively
conjugated in the first and the second antibracket and also $
{\bar{\Phi}}_{A} $
for the bar-variables. The antibracket in equation
(\ref{mastereq}) is defined by
\begin{eqnarray}
(F,G) = (F,G)_{1} + (F,G)_{2} & = &
\frac{\stackrel{\leftarrow}{\delta}F}{\delta \Phi^{A}}
\frac{\stackrel{\rightarrow}{\delta}G}{\delta \Phi^{*}_{A(1)}}
-
\frac{\stackrel{\leftarrow}{\delta}F}{\delta \Phi^{*}_{A(1)}}
\frac{\stackrel{\rightarrow}{\delta}G}{\delta \Phi^{A}}
\nonumber \\
& & + \frac{\stackrel{\leftarrow}{\delta}F}{\delta \Phi^{A}}
\frac{\stackrel{\rightarrow}{\delta}G}{\delta \Phi^{*}_{A(2)}}
-
\frac{\stackrel{\leftarrow}{\delta}F}{\delta \Phi^{*}_{A(2)}}
\frac{\stackrel{\rightarrow}{\delta}G}{\delta \Phi^{A}}.
\end{eqnarray}
Here $ V $ acts only on the bar-variables and is defined as
\begin{equation}
V = V_{1} + V_{2} = \Phi^{*}_{A(2)}
\frac{\stackrel{\rightarrow}{\delta}}{\delta{\bar{\Phi}}_{A}}
- \Phi^{*}_{A(1)} \frac{\stackrel{\rightarrow}{\delta}}{\delta
{\bar{\Phi}}_{A}} .
\end{equation}
The requirement $ bingh(S)=(0,0) $
implies that the equation (\ref{mastereq}) splits into two
parts :
\begin{equation}
{1\over2}(S,S)_{1} + V_{1}S =0= {1\over2}(S,S)_{2} + V_{2}S.
\label{doublemastereq}
\end{equation}
The resolution of (\ref{doublemastereq}) is performed along the
lines of homological
perturbation theory ; one develops $ S $ with respect to the
resolution degree
\begin{equation}
S=\sum_{k=0}^{\infty} {\raise1.2pt\hbox{$ \stackrel{(k)}{S}$}},
\end{equation}
where the boundary terms of $ {\raise1.2pt\hbox{$
\stackrel{(1)}{S}$}} $,
${\raise1.2pt\hbox{$ \stackrel{(2)}{S}$}}$ and
${\raise1.2pt\hbox{$ \stackrel{(3)}{S}$}}$ have been given in
the
preceeding subsection.
Note that the resolution of the master equation
will also fix all the remaining terms of the two Koszul-Tate
operators in
such a way that they become nilpotent and anticommuting off-
shell (this
follows from the generalized Jacobi identity).
The equation in resolution degree 0 is satisfied if
${\raise1.2pt\hbox{$ \stackrel{(1)}{S}$}}$
contains only the already given boundary terms :
\begin{equation}
{\raise1.2pt\hbox{$ \stackrel{(1)}{S}$}}= \stackrel{(-
1,0)*}{q_{i(1)}}R^{i}_{\alpha_{0}}\stackrel{(1,0)}{
\varphi^{\alpha_{0}}} +\stackrel{(0,-
1)*}{q_{i(1)}}R^{i}_{\alpha_{0}}\stackrel{(0,1)}{
\varphi^{\alpha_{0}}}.
\end{equation}

\pagebreak

\noindent
The equation in resolution degree 1 is written as (see
\cite{FisHen1})
\begin{equation}
({\raise1.2pt\hbox{$ \stackrel{(0)}{S}$}},{\raise1.2pt\hbox{$
\stackrel{(2)}{S}$}})^{(0)}+
({\raise1.2pt\hbox{$ \stackrel{(1)}{S}$}},{\raise1.2pt\hbox{$
\stackrel{(2)}{S}$}})^{(1)}+
V{\raise1.2pt\hbox{$ \stackrel{(2)}{S}$}} +
{1\over2}({\raise1.2pt\hbox{$
\stackrel{(1)}{S}$}},{\raise1.2pt\hbox{$
\stackrel{(1)}{S}$}})^{(0)}
=0 \label{masterres1}
\end{equation}
where ({\em i}) $ (\cdot,\cdot)^{(k)} $ stands for the pieces
of the antibrackets
containing only the antifields of resolution degree $ (k+1) $
and ({\em ii}) the
last term vanishes due to the abelian structure of the model.
Note that the equation
(\ref{masterres1}) also splits into
two pieces according to the new ghost bigrading. One can check
that in addition to
the boundary terms given in (\ref{S2}) the
only supplementary terms contained in ${\raise1.2pt\hbox{$
\stackrel{(2)}{S}$}}$ are
\begin{eqnarray}
-{1\over2}\stackrel{(-1,0)*}{q_{i(1)}}\stackrel{(-
1,0)*}{q_{j(1)}}T^{ij}_{\alpha_{1}}
\stackrel{(2,0)}{\varphi^{\alpha_{1}}}-
{1\over2}\stackrel{(-1,0)*}{q_{i(1)}}\stackrel{(0,-
1)*}{q_{j(2)}}T^{ij}_{\alpha_{1}}
\stackrel{(1,1)}{\varphi^{\alpha_{1}}} \nonumber \\
-{1\over2}\stackrel{(0,-1)*}{q_{i(2)}}\stackrel{(0,-
1)*}{q_{j(2)}}T^{ij}_{\alpha_{1}}
\stackrel{(0,2)}{\varphi^{\alpha_{1}}}.
\end{eqnarray}
These terms are due to the fact that we only have reducibility
on-shell. The
equation at resolution degree 2 reads
\begin{equation}
\sum_{k=0}^{2}({\raise1.2pt\hbox{$
\stackrel{(k)}{S}$}}),{\raise1.2pt\hbox{$
\stackrel{(3)}{S}$}})^{(k)}
+ V{\raise1.2pt\hbox{$ \stackrel{(3)}{S}$}} +
({\raise1.2pt\hbox{$ \stackrel{(2)}{S}$}},
{\raise1.2pt\hbox{$ \stackrel{(1)}{S}$}})^{(0)}+
{1\over2}({\raise1.2pt\hbox{$
\stackrel{(2)}{S}$}},{\raise1.2pt\hbox{$
\stackrel{(2)}{S}$}})^{(1)} =0.
\end{equation}
Again the last two terms vanish because the model is abelian
and because the
$ T^{ij}_{\alpha_{1}} $ do not depend on the fields.
A close inspection of the two resulting equations shows that
out of the
50 {\em a priori} possible supplementary terms, only the
following are needed to
satisfy the master equations at this level :
\begin{equation}
{1\over 2}\stackrel{(-1,0)*}{q_{i(1)}}\stackrel{(-1,-
1)}{{\bar{q}}_{j}}T^{ij}_{\alpha_{1}}
\stackrel{(2,1)}{\pi^{\alpha_{1}}}+
{1\over 2}\stackrel{(0,-1)*}{q_{i(2)}}\stackrel{(-1,-
1)}{\bar{q}_{j}}T^{ij}_{\alpha_{1}}
\stackrel{(1,2)}{\pi^{\alpha_{1}}}.
\end{equation}
One can then verify that it is possible to choose
${\raise1.2pt\hbox{$ \stackrel{(k)}{S}$}} = 0$
for $ k>3 $. This completes our derivation of the solution of
the extended master
equation.
Using the identifications (\ref{dic1}),(\ref{dic2}) and
(\ref{dic3}), we get for the
Freedman-Townsend model
\pagebreak

\begin{eqnarray}
S & = & S_{0} + \int d^{4}x \; \Bigg\{ {\raise.1pt\hbox{$
\stackrel{(-1,0)*}{B^{\mu\nu}_{(1)}}$}}
\nabla_{\mu}\stackrel{(1,0)}{\varphi_{\nu}} +
{\raise.1pt\hbox{$ \stackrel{(0,-1)*}{B^{\mu\nu}_{(2)}}$}}
\nabla_{\mu}\stackrel{(0,1)}{\varphi_{\nu}} \nonumber \\
& & + {\raise.1pt\hbox{$ \stackrel{(-1,-
1)*}{\bar{B}^{\mu\nu}_{(1)}}$}}
\nabla_{\mu}\stackrel{(1,1)}{\pi_{\nu}} + (\stackrel{(-1,-
1)*}{\varphi^{\mu}_{(2)}}
- \stackrel{(-1,-
1)*}{\varphi^{\mu}_{(1)}})\stackrel{(1,1)}{\pi_{\mu}} \nonumber
\\
& & + \stackrel{(-2,0)*}{\varphi^{\mu}_{(1)}}\nabla_{\mu}
\stackrel{(2,0)}{\varphi} + {1 \over 2}(\stackrel{(-1,-
1)*}{\varphi^{\mu}_{(1)}} +
\stackrel{(-1,-
1)*}{\varphi^{\mu}_{(2)}})\nabla_{\mu}\stackrel{(1,1)}{\varphi}
+
\stackrel{(0,-
2)*}{\varphi^{\mu}_{(2)}}\nabla_{\mu}\stackrel{(0,2)}{\varphi}
\nonumber \\
& & - {1\over 4} [{\raise.1pt\hbox{$ \stackrel{(-
1,0)*}{B^{\mu\nu}_{(1)}}$}},
{\raise.1pt\hbox{$ \stackrel{(-
1,0)*}{B^{\rho\sigma}_{(1)}}$}}]\epsilon_{\mu\nu\rho\sigma}
\stackrel{(2,0)}{\varphi}
- {1\over 4} [{\raise.1pt\hbox{$ \stackrel{(-
1,0)*}{B^{\mu\nu}_{(1)}}$}},
{\raise.1pt\hbox{$ \stackrel{(0,-
1)*}{B^{\rho\sigma}_{(2)}}$}}]\epsilon_{\mu\nu\rho\sigma}
\stackrel{(1,1)}{\varphi}
- {1\over 4} [{\raise.1pt\hbox{$ \stackrel{(0,-
1)*}{B^{\mu\nu}_{(2)}}$}},
{\raise.1pt\hbox{$ \stackrel{(0,-
1)*}{B^{\rho\sigma}_{(1)}}$}}]\epsilon_{\mu\nu\rho\sigma}
\stackrel{(0,2)}{\varphi} \nonumber \\
& &+ \stackrel{(-2,-
1)*}{\bar{\varphi}^{\mu}_{(1)}}\nabla_{\mu}\stackrel{(2,1)}{\pi
}
- {1 \over 2}\stackrel{(-2,-
1)*}{\pi^{\mu}_{(1)}}\nabla_{\mu}\stackrel{(2,1)}{\pi}
+ (\stackrel{(-2,-1)*}{\varphi_{(2)}} - \stackrel{(-2,-
1)*}{\varphi_{(1)}})
\stackrel{(2,1)}{\pi} \nonumber \\
& & - \stackrel{(-1,-
2)*}{\bar{\varphi}^{\mu}_{(1)}}\nabla_{\mu}\stackrel{(1,2)}{\pi
}
- {1 \over 2}\stackrel{(-1,-
2)*}{\pi^{\mu}_{(2)}}\nabla_{\mu}\stackrel{(1,2)}{\pi}
+ (\stackrel{(-1,-2)*}{\varphi_{(2)}} - \stackrel{(-1,-
2)*}{\varphi_{(1)}})
\stackrel{(1,2)}{\pi} \nonumber \\
& & + {1 \over 4} [{\raise.1pt\hbox{$ \stackrel{(-
1,0)*}{B^{\mu\nu}_{(1)}}$}},
{\raise.1pt\hbox{$ \stackrel{(-1,-
1)}{\bar{B}^{\rho\sigma}}$}}]\epsilon_{\mu\nu\rho\sigma}
\stackrel{(2,1)}{\pi}
+ {1\over 4} [{\raise.1pt\hbox{$ \stackrel{(0,-
1)*}{B^{\mu\nu}_{(2)}}$}},
{\raise.1pt\hbox{$ \stackrel{(-1,-
1)}{\bar{B}^{\rho\sigma}}$}}]\epsilon_{\mu\nu\rho\sigma}
\stackrel{(1,2)}{\pi} \Bigg\}.
\end{eqnarray}

This solution is by construction invariant under the two
following nilpotent and anticommuting
transformations :
\begin{equation}
s_{1}=(S,\cdot)_{1} + V_{1} \qquad {\rm and} \qquad
s_{2}=(S,\cdot)_{2} + V_{2}.
\label{s1s2}
\end{equation}
as well as under any linear combinations of these
transformations.

\section{Gauge fixing}
\subsection{Equivalence with the ordinary BRST formalism}

The following change of variables
will be usefull : $ u^{*}_{A} =
{1\over2}(\Phi^{*}_{A(2)}+\Phi^{*}_{A(1)}) $ and
$ v^{*}_{A} = {1\over2}(\Phi^{*}_{A(2)}-\Phi^{*}_{A(1)})
$.According to reference \cite{Hen}, the gauge fixed action is
given by
\begin{equation}
\exp iS_{gf}=\exp([\hat{K},\bar{\Delta}]) \exp iS
\big\vert_{antifields=0}. \label{GFaction}
\end{equation}
Here we make the choice $ \hat{K}= -\Psi(\Phi^{A}) $ in order
to make the
comparison with the ordinary BRST treatment. The gauge fixed
action in
(\ref{GFaction}) then simply becomes (see \cite{Hen}) :
\begin{equation}
S_{gf}=S(\Phi^{A},u^{*}_{A}=\frac{\stackrel{\leftarrow}{\delta}
\Psi}{\delta\Phi^{A}},
v^{*}_{A}=0,{\bar{\Phi}}_{A}=0).
\end{equation}
When expliciting this last expression, one obtains
\begin{eqnarray}
S_{gf} & = & S_{0}+\frac{\stackrel{\leftarrow}{\delta}\Psi}{\delta
q^{i}}R^{i}_{\alpha_{0}} \phi^{\alpha_{0}}-
\frac{\stackrel{\leftarrow}{\delta}\Psi}{\delta\bar{\phi}^{
\alpha_{0}}}
\pi^{\alpha_{0}} \nonumber \\
& &
+\frac{\stackrel{\leftarrow}{\delta}\Psi}{\delta\phi^{\alpha_{0
}}}
Z^{\alpha_{0}}_{\alpha_{1}}\phi^{\alpha_{1}}+
{1\over
2}\frac{\stackrel{\leftarrow}{\delta}\Psi}{\delta\bar{\phi}^{\a
lpha_{0}}}
Z^{\alpha_{0}}_{\alpha_{1}}\tilde{\phi}^{\alpha_{1}}-
{1\over2}\frac{\stackrel{\leftarrow}{\delta}\Psi}{\delta q^{i}}
\frac{\stackrel{\leftarrow}{\delta}\Psi}{\delta
q^{j}}T^{ij}_{\alpha_{1}}
 \phi^{\alpha_{1}} \nonumber \\
& & -{1\over2}\frac{\stackrel{\leftarrow}{\delta}\Psi}{\delta
\pi^{\alpha_{0}}}
Z^{\alpha_{0}}_{\alpha_{1}}\bar{\pi}^{\alpha_{1}}-
\frac{\stackrel{\leftarrow}{\delta}\Psi}{\delta
{\bar{\phi}}^{\alpha_{1}}}
\tilde{\pi}^{\alpha_{1}} -
\frac{\stackrel{\leftarrow}{\delta}\Psi}{\delta
{\tilde{\phi}}^{\alpha_{1}}}
\bar{\pi}^{\alpha_{1}},
\end{eqnarray}
where we have made the following change of variables
\begin{equation}
(\stackrel{(1,0)}{\varphi^{\alpha_{0}}}+\stackrel{(0,1)}{
\varphi^{\alpha_{0}}})= \phi^{\alpha_{0}} \qquad {\rm and} \qquad
(\stackrel{(1,0)}{\varphi^{\alpha_{0}}}-
\stackrel{(0,1)}{\varphi^{\alpha_{0}}})= \bar{\phi}^{\alpha_{0}},
\end{equation}
\begin{equation}
(\stackrel{(2,0)}{\varphi^{\alpha_{1}}}+\stackrel{(1,1)}{
\varphi^{\alpha_{1}}}
+\stackrel{(0,2)}{\varphi^{\alpha_{1}}})=\phi^{\alpha_{1}},
\end{equation}
\begin{equation}
(\stackrel{(2,0)}{\varphi^{\alpha_{1}}}-
\stackrel{(1,1)}{\varphi^{\alpha_{1}}}
+\stackrel{(0,2)}{\varphi^{\alpha_{1}}})
=\bar{\phi}^{\alpha_{1}},
\end{equation}
\begin{equation}(\stackrel{(2,0)}{\varphi^{\alpha_{1}}}-
\stackrel{(0,2)}{\varphi^{\alpha_{1}}})
= {1 \over 2} \tilde{\phi}^{\alpha_{1}},
\end{equation}
\begin{equation}
\stackrel{(1,1)}{\pi^{\alpha_{0}}}= -{1 \over
2}\pi^{\alpha_{0}},
\end{equation}
\begin{equation}
(\stackrel{(2,1)}{\pi^{\alpha_{1}}}-
\stackrel{(1,2)}{\pi^{\alpha_{1}}})= -{1 \over
2}\tilde{\pi}^{\alpha_{1}} \qquad {\rm and} \qquad
(\stackrel{(2,1)}{\pi^{\alpha_{1}}}+\stackrel{(1,2)}{\pi^{
\alpha_{1}}})= -{1 \over 2}\bar{\pi}^{\alpha_{1}}.
\end{equation}

{}From the transformations (\ref{s1s2}), one can define their
gauge fixed counterparts,
respectively denoted by $ s_{1\ gf} $ and $ s_{2\ gf} $ and
obtained by first
calculating $ s_{1} $ and $ s_{2} $ and fixing the gauge
afterwards.
Let us define $ s=s_{1\ gf}+s_{2\ gf} $. The
gauge fixed action $ S_{gf} $ is then invariant under $ s $
explicitly defined by :
\begin{eqnarray}
sq^{i}=R^{i}_{\alpha_{0}}\phi^{\alpha_{0}} -
\frac{\stackrel{\leftarrow}{\delta}\Psi}{\delta
q^{j}}T^{ij}_{\alpha_{1}}\phi^{\alpha_{1}}
&\qquad
s\phi^{\alpha_{0}}=Z^{\alpha_{0}}_{\alpha_{1}}\phi^{\alpha_{1}}
\nonumber \\
s\bar{\phi}^{\alpha_{0}}={1 \over
2}Z^{\alpha_{0}}_{\alpha_{1}}\tilde{\phi}^{\alpha_{1}}-
\pi^{\alpha_{0}}
&\qquad
s\pi^{\alpha_{0}}=-{1 \over
2}Z^{\alpha_{0}}_{\alpha_{1}}\bar{\pi}^{\alpha_{1}}\nonumber \\
s\bar{\phi}^{\alpha_{1}}=-\tilde{\pi}^{\alpha_{1}}
&\qquad
s\tilde{\phi}^{\alpha_{1}}=-\bar{\pi}^{\alpha_{1}}\nonumber \\
s\phi^{\alpha_{1}}=s\tilde{\pi}^{\alpha_{1}}=s\bar{\pi}^{\alpha
_{1}}=0.
\end{eqnarray}

As expected on general grounds, one can check that $ s $ is
nilpotent on the stationary surface defined by the equations
of motion associated to the gauge fixed action. In fact $s$ is
nilpotent off-shell for all
generators with the exception of $ s^2 q^i =
-\stackrel{\leftarrow}{\delta} S_{gf} / \delta q^{j}
T^{ij}_{\alpha_{1}}\phi^{\alpha_{1}}$.
However, the choice $ \hat{K}= -\Psi(\Phi^{A}) $ spoils the
invariance of $S_{gf}$ under
$ s_{1\ gf} $ and $ s_{2\ gf} $ separately.

Translating these results to the Freedman-Townsend model with
the following choice of
the gauge fixing fermion
\begin{equation}
\Psi= - \int d^4x\; \Bigg\{\alpha(\nabla^\mu\bar\phi^\nu)
B_{\mu\nu}+\beta(\phi_\mu\nabla^\mu\bar\phi-
\bar\phi^{\mu}(\pi_{\mu} - {1 \over 2} \nabla_\mu\tilde\phi)
+\gamma\bar\pi\bar\phi +
\delta\tilde\pi\tilde\phi \Bigg\}
\end{equation}
gives
\begin{eqnarray}
S_{gf} & = & S_0 + \int d^4x \Bigg\{ -\alpha
(\nabla^{[\mu}\bar\phi^{\nu]})\nabla_\mu\phi_\nu -\beta
\pi^{\mu}\pi_{\mu} - \alpha (\nabla^\mu B_{\mu\nu})\pi^\nu
\nonumber \\
& & - \beta (\nabla_\mu\bar\phi)\nabla^\mu\phi +{\alpha \over
2} (\nabla^\mu B_{\mu\nu})\nabla^{\nu}
\tilde\phi + {\beta \over 4} \nabla_\mu\tilde\phi
\nabla^\mu\tilde\phi
+ {{\alpha^{2}} \over 4}[ \nabla^\mu\bar\phi^\nu, \nabla^\rho
\bar\phi^\sigma ] \epsilon_{\mu\nu\rho\sigma}
\phi \nonumber \\
& & + \beta (\nabla_\mu\bar\phi^\mu )\bar\pi - \beta
(\nabla^\mu\phi_\mu )\tilde\pi + (\gamma - \delta)
\bar\pi\tilde\pi \Bigg\} \label{SBRST}
\end{eqnarray}
This choice of $\Psi$ gives a well defined path integral as can
be seen by elimination of the
auxiliary fields $\pi_\mu, \bar\pi, \tilde\pi$, and will allow
us the comparision with the
BRST-anti-BRST gauge fixed action given below.
The gauge fixed BRST symmetry is given by :
\begin{eqnarray}
s B_{\mu\nu}^a  =  \nabla_{[\mu}\phi_{\nu]} + {1\over
2}(\nabla^\rho\bar\phi^\sigma)^b f^{cab}
\epsilon_{\mu\nu\rho\sigma}\phi^a \nonumber \\
s \phi_\mu =\nabla_\mu\phi\qquad s \bar\phi_\mu = {1 \over 2}
\nabla_\mu\tilde\phi-\pi_\mu\qquad
s \pi_\mu = - {1 \over 2} \nabla_\mu\bar\pi \nonumber \\
s \bar\phi = -\tilde\pi\qquad s \tilde\phi = -\bar\pi\qquad s
A_\mu =s \phi =s \bar\pi =s
\tilde\pi = 0.
\end{eqnarray}

Apart from the cohomologically trivial variables
$ \bar{\phi}^{\alpha_{0}}, B^{\alpha_{0}},
\bar{\phi}^{\alpha_{1}}, \tilde{\phi}^{\alpha_{1}},
\tilde{B}^{\alpha_{1}}, \bar{B}^{\alpha_{1}} $ and their
antifields,
this result, with the identifications (\ref{dic1}),
(\ref{dic2}) and (\ref{dic3}), coincides with the result
obtained from the usual BRST formalism
\cite{AlvGri,BatGom,BauBer} where some of the above variables
are reintroduced as a non-minimal sector for
gauge fixing purposes.

\subsection{BRST-anti-BRST gauge fixed action}
According to references \cite{BatLav3,BatLav4}, the BRST-anti-
BRST invariant
gauge fixed action ${\cal S}_{gf}$ \footnote{An alternative
derivation of this gauge fixed
action in the framework of the standard BRST formalism
and a constructive approach to the on-shell nilpotent and
anticommuting gauge fixed BRST-anti-BRST
symmetries from which the above transformations differ by
antisymmetric combinations of the equations
of motion obtained from ${\cal S}_{gf}$ will be given in a
forthcoming paper \cite{GreHen2}.}
is given by :
\begin{eqnarray}
 {\cal S}_{gf} & = & S(\Phi^A,\Phi^{*}_{Ai},\bar{\Phi}^A)
\nonumber \\
 & + & \int d^{4}x \; \Bigg\{ \Phi^{*}_{Ai}\xi^{Ai} +
\left(\bar{\Phi}_{A} -
\frac{\stackrel{\leftarrow}{\delta}F}{\delta\Phi^{A}} \right)
{\lambda^{A}} -
 {1 \over 2} \epsilon_{ij} \xi^{Ai}
\frac{\stackrel{\leftarrow}{\delta}^{2}F}{\delta\Phi^{B}\delta\
Phi^{A}}
\xi^{Bj} \Bigg\}
\end{eqnarray}
where $S(\Phi^A,\Phi^{*}_{Ai},\bar{\Phi}^A)$ is the solution of
the extended master equation
(\ref{mastereq}), $\epsilon_{ij}\quad i,j=1,2$ is completely
antisymmetric with
$\epsilon_{12}=-1$ and $F$ is a bosonic gauge fixing functional
depending on the
$\Phi^A$'s only.
The $\Phi^{*}_{Ai},\bar{\Phi}^A,\xi^{Ai}$ and $\lambda^{A}$'s
play the role of ordinary variables
in the gauge fixed action and must be integrated over in the
path integral just like the $\Phi^A$'s.
The gauge fixed action is invariant under the following two
transformations :
\begin{equation}
s^i\Phi^{A}=\xi^{Ai} \qquad
s^i\Phi^{*}_{Aj} = \delta^i_j{\stackrel{\leftarrow}{\delta}S
\over \delta
\Phi^{A}}(-)^{\epsilon_A} \qquad s^i\lambda^{A}=0
\end{equation}
\begin{equation}
s^i\bar{\Phi}_{A}=\epsilon^{ij}\Phi^{*}_{Aj}(-)^{\epsilon_A+1}
\qquad
s^i\xi^{Aj}=\lambda^{A}\epsilon^{ij}
\end{equation}
where $s^i$ acts as an odd right derivative.

For the Freedmann-Townsend model, the final form of the gauge
fixed action, with the choice
\begin{equation}
F=\int d^4x\;\Bigg\{ {\alpha\over 2}
B^{\mu\nu}B_{\mu\nu}+\beta\phi_\mu\bar\phi^\mu+
\gamma\phi\bar\phi+{\delta\over 2}\tilde\phi\tilde\phi\Bigg\}
\end{equation}
and after elimination of some auxiliary fields (and the
corresponding modifications of the symmetries
(see \cite{Hen2})) is given by \footnote{Here an obvious
renaming of the fields has been performed
in order to make the comparision to the previous result more
transparent :
the {\em new ghost bigrading} has been dropped and the
numerotation of the ghosts has been replaced
by the superscript. The bar now means antighost; no confusion
with the previous bar-variables can
arise, because those variables have been eliminated.} :
\begin{eqnarray}
{\cal S}_{gf} & = & S_0 +\int d^4x\; \Bigg\{ B_{(1)}^{\mu\nu
*}\nabla_\mu\phi_\nu +
B_{(2)}^{\mu\nu *}\nabla_\mu\bar\phi_\nu +\alpha
B^{\mu\nu}\nabla_\mu\pi_\nu \nonumber \\
 & & - {1\over 4} [B_{(1)}^{\mu\nu *},B_{(1)}^{\rho\sigma
*}]\epsilon_{\mu\nu\rho\sigma}\phi
- {1\over 4} [B_{(1)}^{\mu\nu *},B_{(2)}^{\rho\sigma
*}]\epsilon_{\mu\nu\rho\sigma}\tilde\phi
- {1\over 4} [B_{(2)}^{\mu\nu *},B_{(2)}^{\rho\sigma
*}]\epsilon_{\mu\nu\rho\sigma}\bar\phi
\nonumber \\
& & + {\alpha \over 4} [B_{(1)}^{\mu\nu *},B^{\rho\sigma
}]\epsilon_{\mu\nu\rho\sigma}\bar\pi
+ {\alpha \over 4} [B_{(2)}^{\mu\nu *},B^{\rho\sigma
}]\epsilon_{\mu\nu\rho\sigma}\tilde\pi
+ B_{(1)}^{\mu\nu *}\xi^{(1)}_{\mu\nu} + B_{(2)}^{\mu\nu
*}\xi^{(2)}_{\mu\nu} + \alpha\xi^{(1)\mu\nu}\xi^{(2)}_{\mu\nu}
\nonumber \\
& & +\beta (-\pi^\mu\pi_\mu + {1\over
4}(\nabla^\mu\tilde\phi)\nabla_\mu\tilde\phi
-(\nabla^\mu\bar\phi)\nabla_\mu\phi
+ (\nabla^\mu\bar\phi_\mu)\bar\pi -
(\nabla^\mu\phi_\mu)\tilde\pi) \nonumber \\
& & +(\gamma -\delta)\bar\pi\tilde\pi \Bigg\}. \label{Action}
\end{eqnarray}
In a perturbative approach to quantization, this action is
gauge fixed because one treats
the cubic terms as interactions and one can then eliminate the
fields
$B_{(i)}^{\mu\nu *}$ and $\xi^{(i)}_{\mu\nu}$ getting the usual
kinetic term
for the ghosts and antighosts of first order. The integration
over $\pi_\mu, \bar\pi $ and $ \tilde\pi $
then leads to well-defined
propagators for the remaining fields.
The action (\ref{Action}) is invariant under the
transformations :
\begin{eqnarray}
&s^i A_{\mu}=0\qquad s^i B_{\mu\nu}=\xi^{(i)}_{\mu\nu}\qquad
s^i\bar\pi=s^i\tilde\pi=0
\nonumber \\
&s^1\phi_\mu = -\nabla_\mu\phi\qquad s^2\phi_\mu = -\pi_\mu-
{1\over 2}
\nabla_\mu\tilde\phi\qquad
s^1\bar\phi_\mu =\pi_\mu-{1\over 2}
\nabla_\mu\tilde\phi\qquad s^2\bar\phi_\mu =-\nabla_\mu\bar\phi
\nonumber \\
& s^1\phi=0\qquad s^2\phi=-\bar\pi\qquad s^1\tilde\phi=\bar\pi
\qquad s^2\tilde\phi=-\tilde\pi\qquad
s^1\bar\phi=\tilde\pi\qquad
s^2\bar\phi=0
\nonumber \\
&s^1\pi_\mu={1\over 2}\nabla_\mu\bar\pi\qquad
s^2\pi_\mu={1\over 2}\nabla_\mu\tilde\pi
\nonumber \\
&s^i\xi^{(j)a}_{\mu\nu}=-\left( (\nabla_\mu\pi_\nu)^a +
\frac{1}{4} B_{(1)b}^{\rho\sigma *}
f^{cba}\epsilon_{\mu\nu\rho\sigma}\bar\pi_c + \frac{1}{4}
B_{(2)b}^{\rho\sigma *}
f^{cba}\epsilon_{\mu\nu\rho\sigma}\tilde\pi_c
\right)\epsilon^{ij}
\nonumber \\
&s^i B_{(j)}^{\mu\nu *}=\delta^i_j \frac{1}{4}
\epsilon^{\mu\nu\rho\sigma}F_{\rho\sigma}.
\end{eqnarray}

\section{Conclusion}
Even though in this model the fields
$B_i^{\mu\nu *}$ and $\xi^{i}_{\mu\nu}$ are auxiliary, their
elimination is cumbersome
because the matrix
of their quadratic part depends on the second generation ghosts
and the structure constants.
In the abelian case however, this elimination can easily be
performed and the resulting gauge fixed action
is :
\begin{eqnarray}
{\cal S}_{gf} & = & S_0 +\int d^4x\; \Bigg\{
-\alpha(\partial^{[\mu}\bar\phi^{\nu]})\partial_\mu\phi_\nu
+\alpha B^{\mu\nu}\partial_\mu\pi_\nu +\beta (-\pi^\mu\pi_\mu +
{1\over 4}(\partial^\mu\tilde\phi)\partial_\mu\tilde\phi
\nonumber \\
& - & (\partial^\mu\bar\phi)\partial_\mu\phi
+ (\partial^\mu\bar\phi_\mu)\bar\pi -
(\partial^\mu\phi_\mu)\tilde\pi)+(\gamma
-\delta)\bar\pi\tilde\pi
\Bigg\}.
\end{eqnarray}
This result coincides with the corresponding result of equation
(\ref{SBRST})
and the anti-BRST symmetry can be implemented on the same set
of fields than the one used
for the solution of the standard master equation (with an
appropriate non-minimal sector).
Furthermore, because of the linear dependence of the solution
of the master equation
(and the extended master equation) in the antifields, the BRST-
anti-BRST gauge fixed
action is given by
\begin{equation}
{\cal S}_{gf}=S_0 - s^1s^2 F
\end{equation}
and coincides with the result given in reference \cite{BauThi}.
Consequently, it is only for the abelian case that the anti-
BRST symmetry can be implemented
in the same way than for Yang-Mills theory
\cite{BatLav3,BauThi}. In the non-abelian case,
the incorporation of the anti-BRST symmetry needs the
supplementary auxiliary fields
$B_i^{\mu\nu *}$ and $\xi^{i}_{\mu\nu}$.
This is yet another illustration of the general idea underlying
the BRST construction,
namely that a more symmetric description requires in general
more variables.
\section{Acknowledgments}
The authors wish to thank Marc Henneaux for useful discussions.
One of us (R.C.) wishes to thank Professors Radu Balescu and
Marc Henneaux for
their hospitality at ULB. He also acknowledges financial
support from the
European Community, project CEE-Tempus JEP 2814-91/1.

\vfill
\pagebreak

\end{document}